\documentclass[twocolumn,floatfix,showpacs,superscriptaddress]{revtex4}
\usepackage{graphicx}
\usepackage{amsmath}
\usepackage{amssymb}
\usepackage{bm}
\usepackage{color}

\begin{document}

\title{Magnetization signatures of light-induced quantum Hall edge states}
\author{Jan P. Dahlhaus}
\affiliation{Department of Physics, University of California, 
Berkeley, CA 95720, USA}
\author{Benjamin M. Fregoso}
\affiliation{Department of Physics, University of California, 
Berkeley, CA 95720, USA}
\author{Joel E. Moore}
\affiliation{Department of Physics, University of California, 
Berkeley, CA 95720, USA}
\affiliation{Materials Science Division, Lawrence Berkeley National Laboratory, 
Berkeley, CA 95720, USA}
\begin{abstract}
Circularly polarised light opens a gap in the Dirac spectrum of graphene and topological insulator (TI) surfaces, thereby inducing a quantum Hall-like phase. We propose to detect the accompanying edge states and their current by the magnetic field they produce.
The topological nature of the edge states is reflected in the mean orbital magnetization of the sample, which shows a universal linear dependence as a function of a generalized chemical potential -- independent of the driving details and the properties of the material.
The proposed protocol overcomes several typically encountered problems in the realization and measurement of Floquet phases, including the destructive effects of phonons and coupled electron baths and provides a way to occupy the induced edge states selectively.
We estimate practical experimental parameters and conclude that the magnetization signature of the Floquet topological phase may be detectable with current techniques.
\end{abstract}
\maketitle 

The last decade has seen huge steps in the quest to understand \cite{Has10,Qi11,Bee12,Ali12}, produce and detect topological band insulators and superconductors~\cite{Kon07,Hsi08,Xia09,Mou12}. While a variety of materials that realize topological phases are known by now, experimentalists still fight with material-specific imperfections~\cite{And13}. It was proposed though that subjecting non-topological materials to light could provide another, fundamentally different way to induce  topological properties in electronic behaviour~\cite{Oka09,Ino10,Lin11}.

Topological phases induced in this way form part of a wider class of systems termed Floquet topological insulators~\cite{Kit10,Kit10b,Cay13,Gom13,Rud13}. They are intrinsically non-equilibrium systems inheriting their properties from both the oscillating electromagnetic field and the initial band structure of the material.
For example circularly-polarized light is expected to act similar as a static magnetic field on graphene or a TI surface -- opening a gap in the Dirac-spectrum and driving the system in a quantum Hall-like state~\cite{Oka09,Sen14,Kit11,Cal11,Fre13}. Recently experiments have observed first signatures of this effect in the band structure~\cite{Ged13,Oni14,Zha14}, but demonstration of the topological nature of the light-generated phase remains an open problem.

The hallmark of TIs is the existence of protected boundary states~\cite{Has10,Qi11,Bee12,Ali12,Ess11}, a fact that extends to light-induced topological phases~\cite{Oka09,Ino10,Lin11,Rud13}. For example, both integer quantum Hall and light-induced quantum Hall-like phases feature a number of unidirectional edge states that carry a constant current around the sample if occupied. An experimental observation of these light-induced edge states would prove the topological nature of a light-induced gap and is therefore highly desirable. 

\begin{figure}[b] 
\centerline{\includegraphics[width=0.90\linewidth]{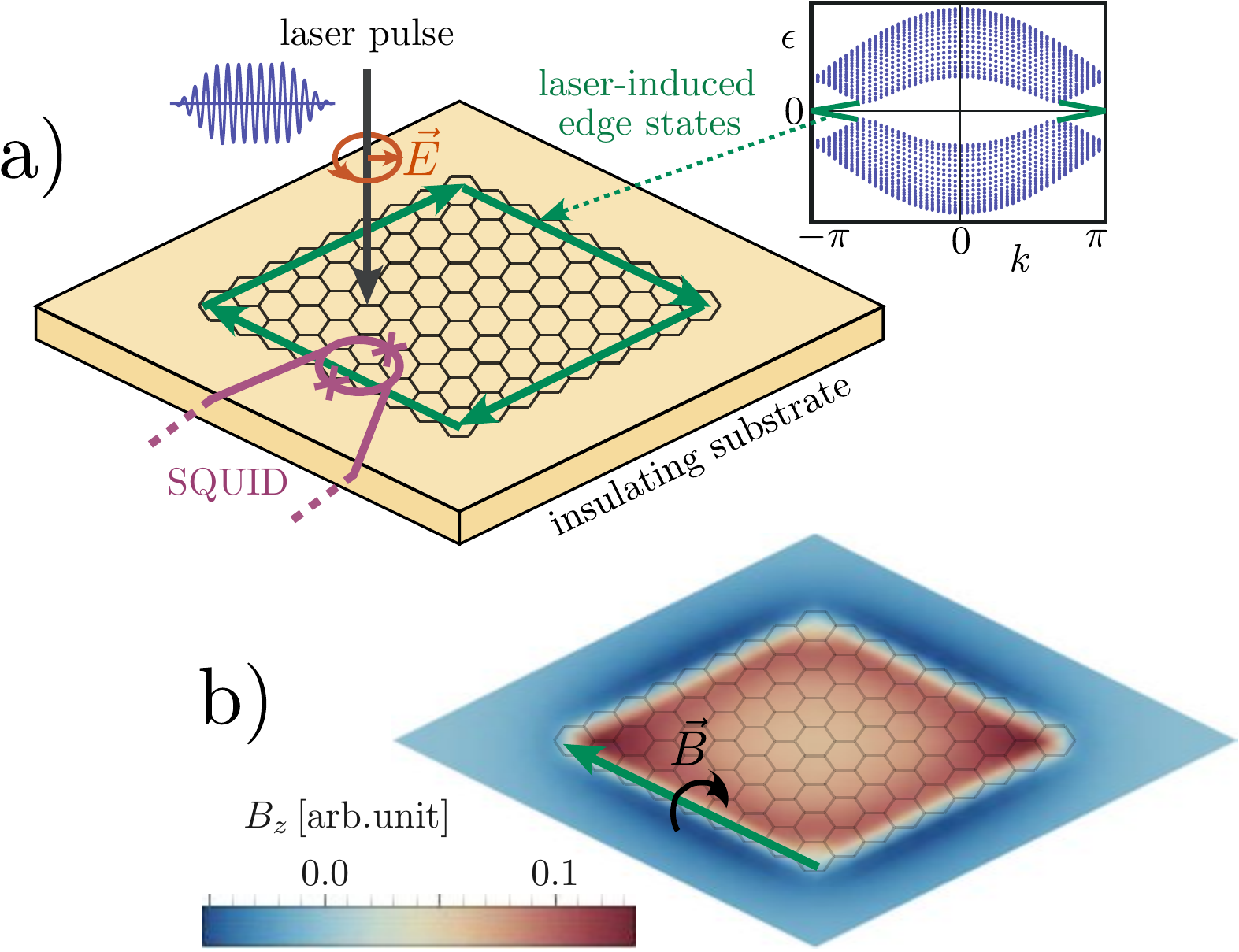}}
\caption{a) Setup and b) Magnetic field pattern produced by a Floquet edge state, see supplementary material. }
\label{fig:Setup}
\end{figure}

In this work we propose an experimental protocol that allows to create such topological edge currents in a controlled fashion and measures them through the magnetic field they produce. The setup we envision is depicted in Fig. \ref{fig:Setup}a). An isolated two-dimensional electronic system, here graphene, is irradiated by circularly-polarized light. Careful design of the laser pulse and control over the initial chemical potential allows to selectively occupy the induced edge states and ensures that phonons do not destroy the effect. 
The resulting edge current produces a magnetic field pattern resembling that of a current loop, see Fig. \ref{fig:Setup}b). We expect that it becomes visible in sensitive measurements of the magnetic field, especially close to the edge of the sample, e.g. by using a SQUID device~\cite{Cla04,Spa14,Now13}. In contrast to all the measurement techniques pursued so far ~\cite{Now13,Syz08,Gu11,Ka11,Fre14}, our approach combines an isolated sample with a non-invasive probing technique and thereby overcomes the destructive effects of coupled leads that are fatal for the driven phase.

{\emph{Hamiltonian and Floquet states}}.
Inspired by recent experiments on TIs \cite{Ged13,Oni14}, we will focus on Dirac-like electron systems and circularly-polarized light to demonstrate our more general concepts. Instead of discussing TIs themselves, we turn to graphene, which also has a Dirac spectrum but is considerably easier to simulate -- however, our results should be transferable. Graphene can be captured by a honeycomb tight-binding Hamiltonian~\cite{Cas09}, 
\begin{align}
H(t)=\sum_{ji}\gamma c^\dag_j c_i+[\bm{E}(t)\cdot \bm{x}_i]\, c^\dag_i c_i
\label{eq:DrivenGraphene}
\end{align}
where the sum runs over nearest neighbours. The second term describes the effect of the rotating electric field $\bm{E}(t)=E_0(\sin \omega t, \cos \omega t)$ caused by the light, where $\bm{x}_i$ denotes the position operator of site $i$. The effects of the much smaller magnetic contribution of the light field can be neglected~\cite{Oka09}. In the following we will use the dimensionless quantity $\mathcal{A}=eE_0 a/\hbar\omega$ to characterise the light intensity and the hopping energy $\gamma= 2\hbar v_F/3a  \approx 2.8$eV as the unit of energy. Here $v_F\approx 10^6$m/s is the Fermi velocity near the Dirac point and $a=1.41\AA$ the carbon-carbon spacing. The simulations are performed on a square array of $L\times L$ graphene unit cells.

In a periodically-driven system like ours, the role of energy (which is not conserved) is taken over by quasienergies $\epsilon_j \in (-\pi/T,\pi/T]$, eigenvalues of the so-called Floquet operator $\mathcal{F}=U(T)$, the time evolution operator $U(t)$ for one period. 
The corresponding eigenstates return to their initial form after propagation for one cycle,  $\mathcal{F} |f_j\rangle =e^{-i\epsilon_j T}|f_j\rangle$, which makes the set of all of these so-called Floquet states a very useful basis.

The circularly-polarized radiation opens a gap in the quasienergy spectrum of graphene, as illustrated in the inset of Fig.~\ref{fig:Setup} for a strip geometry sample~\cite{footnote1}. This gap can be regarded as a gap "of magnetic type" due to the formal similarity to the gap induced by a time-reversal symmetry breaking field \cite{Kit11}. The topological edge states that arise (green) are Floquet eigenstates with quasienergies crossing the bulk quasienergy gap.
They come hand in hand with a non-trivial total winding of the Floquet states in the Brillouin zone~\cite{Kit10,Gom13,Rud13}, in analogy to the corresponding time-independent situation~\cite{Tho82}.

\begin{figure*}[tb] 
\centerline{\includegraphics[width=0.96\linewidth]{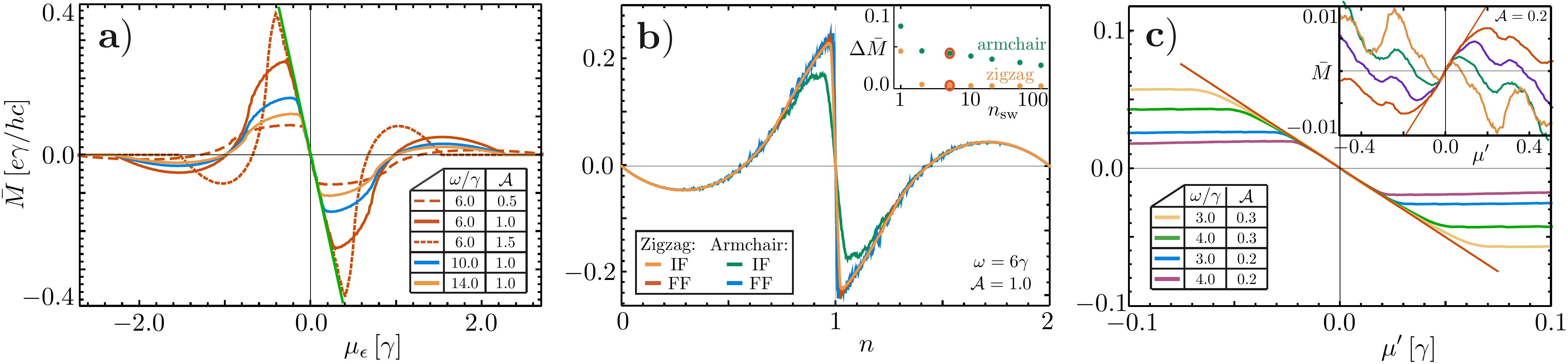}}
\caption{Mean magnetization a) as a function of the quasienergy potential $\mu_\epsilon$. Collapsing curves for various laser parameters demonstrate the universal linear dependence caused by the topological edge states. b) as a function of initial filling (IF) and artificial Floquet filling (FF) for a switch-on process that lasts $n_{\rm sw}\!=\!5$ laser cycles , demonstrating that this short period is good enough to transfer initial energy states mostly into Floquet states.
The inset shows the quality $\Delta M$ of the switch-on process as a function of $n_{\rm sw}$.
c) as a function of the generalized chemical potential $\mu'$ demonstrating the universal linear behaviour of the topological edge states for the experimental protocol. Inset: collapsing curves in a low frequency regime, with frequencies $\omega/ \gamma=0.4,0.5,0.6,0.8$ from red to orange. Main plots were simulated for $L=30$, the insets for $L=20$ (b) and $L=40$ (c) \cite{footnote2}.
}
\label{fig:plots}
\end{figure*}

{\emph{ Orbital magnetization of Floquet states}}. 
The orbital magnetization operator for electrons in an isolated two-dimensional sample of area $V$ is given by
\begin{align}
\hat{M}(t)=-\frac{e}{2\hbar cV}\hat{\bm{r}}\times \hat{\bm{v}}=-\frac{e}{2\hbar cV}\hat{\bm{r}}\times i[\hat{\bm{r}},H(t)],
\end{align}
in Gaussian units. For the electric gauge chosen in Eq.~(\ref{eq:DrivenGraphene}), $\hat{M}(t)$ becomes time-independent.

An electron that is found in a Floquet eigenstate $ |f_i(t)\rangle$ would contribute $M_{ii}(t)= \langle f_i(t)|\hat{M}|f_i(t)\rangle$
to the orbital magnetization of the sample. We will be particularly interested in the mean magnetization $\bar{M}_{ii}$, obtained by averaging $M_{ii}(t)$ over one period of the driving field. Written in terms of the Fourier components $|\phi^{(l)}_i\rangle$ of the periodic part $|\phi_{i}(t)\rangle=|\phi_{i}(t+T)\rangle$ of the Floquet state, $|f_i(t)\rangle= e^{-i\epsilon_i t} |\phi_{i}(t)\rangle$, it becomes
\begin{align}
\bar{M}_{ii}&= \frac{1}{T}\int \limits_{t=0}^{T} dt\;  M_{ii}(t)=\sum_l\langle \phi^{(l)}_i|\hat{M}|\phi^{(l)}_i\rangle.
\label{eq:FloquetMag}
\end{align}

The last expression can be understood as the expectation value of a block diagonal operator $\bar{M}_F$  with block components $(\bar{M}_F)_{kl}=\hat{M}\delta_{kl}$ over the vector of Fourier components $|\Phi_i\rangle=(\ldots,|\phi^{(1)}_i\rangle,|\phi^{(0)}_i\rangle,|\phi^{(-1)}_i\rangle,\ldots)^T$. The $|\Phi_i\rangle$ in turn are the eigenvalue of the so-called Floquet Hamiltonian ~\cite{Shi65}, as discussed in more detail in the supplementary material.

Hence we can interpret the formula for the mean magnetization as the magnetization of a non-driven system with a multi-orbital (corresponding to the frequency components) Hamiltonian. It follows that all statements known about the orbital magnetization in non-driven systems, ~\cite{Res10}, apply correspondingly for the mean magnetization of irradiated samples, including statements about the influence of the Berry curvature and, in particular, the orbital magnetization contribution of topological edge states. This conclusion is a major result of this manuscript.

\emph{Signatures of light-induced quantum Hall edge states.} Now consider an infinite strip of graphene with conserved momentum $k$. The edge states are Floquet states localized at the edges of the sample, with mean velocity $v=\tfrac{d\epsilon_k}{\hbar dk}$~\cite{Gon14,Per14}. 
Let's assume for the moment that we can occupy Floquet states at will, filling e.g. the lower Floquet band as shown in the inset of Fig.~\ref{fig:Setup} from the lowest quasienergy up to a "quasienergy chemical potential" $\mu_\epsilon$. When occupying additional edge Floquet states $d\mu_\epsilon$, the current carried by the edge changes by $dI=-e v \tfrac{dk}{2\pi}=- \tfrac{e}{h}d\mu_\epsilon$. Here $d\mu_\epsilon$ is the analogue of a chemical potential for the quasienergies $\epsilon$. 

We already saw the effects of this current in the local magnetic field pattern in Fig.~\ref{fig:Setup}b). In such a square sample, the edge current circulates around the sample and produces a magnetic field, which, through Maxwells equations can be captured by a magnetization of strength $M=I/c$.
The mean orbital magnetization of a quantum Hall edge state is thus given by the term
\begin{align}
dM_{\rm edge}=-\frac{e}{h c} C \,d\mu_\epsilon, 
\label{eq:Medge}
\end{align}
with $C$ being the number of protected edge states crossing the respective gap, analogous to the non-driven case~\cite{Res10}. This universal linear mean magnetization behaviour is a unique signature of the edge states that depends solely on the topological properties of the system. Our numerical results in Fig. \ref{fig:plots}a) clearly demonstrate this universality through the collapse of the mean magnetization $\bar{M}(\mu_\epsilon)=\sum_{\epsilon_i<\mu_\epsilon}\!\!\bar{M}_{ii}$ onto a single (green) line $\bar{M}=-\frac{e}{h c}\mu_\epsilon$ for various laser parameters.

\emph{Experimental protocol and practical considerations.}

Three major challenges need to be overcome to realize and detect the Floquet phase experimentally: 

The first one is to minimize energy exchange with the environment, which would be fatal for the Floquet states. We ensure this by our choice of a non-invasive measurement, i.e. by detecting the effect via its magnetic field, thereby avoiding direct contact of the sample with fermionic reservoirs. This distinguishes our approach from conventional measurement setups~\cite{Syz08,Gu11,Ka11,Fre14}. 

As a second challenge, we need to ensure that the inherent relaxation processes of the system -- phonons and electronic interactions -- do not influence the dynamics strongly.
This is achieved by choosing a very strong but short laser pulse: not only is the desired non-linear effect induced to a measurable degree -- the strong driving also renders the competing (decay) energy scales less important. Combined with the very short pulse duration, relaxation processes can be kept to a minimum -- they happen on longer time scales. 


Let us now consider how to overcome the third major challenge: we need to ensure that we can selectively occupy the Floquet states. Before the driving is present, the system is in a low temperature ground state with chemical potential $\mu$, where all the electrons are essentially found in energy eigenstates $|e_i\rangle$ with energies $E_i$. According to the adiabatic theorem for periodically-driven systems~\cite{Bre90}, energy eigenstates can be turned into Floquet states by an adiabatic switch-on of the driving strength. We find that the times required for an approximately adiabatic switch-on are surprisingly short. 
 
To be more precise we smoothly turn on the electric field amplitude over a switch-on period $t_{\rm sw}$. The electrons' propagation through this time leads to new states $|\psi_i \rangle=U_{\rm switch} |e_i\rangle$.
After the switch-on process, the Hamiltonian becomes strictly periodic and the time evolution of the electronic states is best described by writing them as a superposition of Floquet eigenstates, $|\psi_i (t)\rangle=\sum_j a_{ij} |f_j(t)\rangle$. The prefactors $a_{ij}=\langle f_j(0)|U_{\rm switch} |e_i\rangle$ depend only on the switch-on process and therefore the shape of the laser pulse. For an adiabatic switch-on, $a_{ij}=\delta_{ij}$.
Together with the choice of initial chemical potential $\mu$, adiabaticity thus allows us to occupy Floquet states in a controlled way. 

The expectation value of the magnetization can be expressed as
\begin{align}
M(t)&=\sum_{E_i<\mu}\sum_{j,k} a_{ij}^*a_{ik}  e^{i(\epsilon_j-\epsilon_k) t} M_{jk}(t),
\end{align}
with $M_{jk}(t)= \langle \phi_j(t)|\hat{M}(t)|\phi_k(t)\rangle$. 
With a perfectly adiabatic switch-on process ($a_{ij}=\delta_{ij}$) the mean magnetization is thus $\bar{M}=\sum_{E_i<\mu}\bar{M}_{ii}$, equal to the situation with artificial quasienergy chemical potential $\mu_\epsilon$. 

The simulation results in Fig.~\ref{fig:plots}b) demonstrate how small a switch-on time is needed to obtain a nearly complete adiabatic transition into Floquet states. To be more precise, we plot the mean magnetization obtained after a switch-on duration of $t_{\rm sw}=5 T$  as a function of initial filling fraction $n$ (IF) and compare it to the ``artificial" magnetization curve as plotted in Fig.~\ref{fig:plots}a), translated into a function of Floquet band filling fraction (FF). The transition from energy to Floquet states works excellently for zigzag termination of the sample, with nearly indistinguishable curves. For armchair edges, the adiabaticity is good in overall, with some deviations closer to the Dirac point. A realistic sample is expected to show a behaviour between the two extremes. The inset of Fig. \ref{fig:plots}b) illustrates the adiabaticity for different switch-on times, measured in terms of the integrated distance between the curves, $\Delta\bar{M}=\int \!dn(\bar{M}_{\rm FF}-\bar{M}_{\rm IF})$.

While $\bar{M}$ as seen in Fig.~\ref{fig:plots}b) depends linearly on $n$ when filling edge states, the gradient is not universal.
The universality uncovered in  Eq.~(\ref{eq:Medge}) can be seen for the laser pulse protocol after transforming $n$ into a generalized chemical potential $d\mu'=L^2 dn /\rho(n)$ via the density of Floquet states $\rho$. In the limit of weak driving this can be done in the edge state region by estimation of $\rho$ through the size of the light-induced gap $\Delta_0\approx\tfrac{2(\hbar v_F/a)^2}{\hbar \omega}\mathcal{A}^2$ and the distance of the Dirac cones in the edge Brillouin zone, $\Delta k\approx 2/\sqrt{3}a$~\cite{Kit11,Fre13}. We thus obtain $d\mu'=(2\pi L/8\Delta_0)\, dn$ and as shown in Fig. \ref{fig:plots}c) the universal linear dependence is recovered, providing a direct way to see the topological nature of the edge states.


Let us turn to lower laser frequencies now. In this regime, the quasienergy spectrum becomes folded and adiabaticity is not able to ensure the filling of Floquet states in the right order. Nevertheless, varying $\mu$ closely around the Dirac point still occupies the Floquet edge states in a controlled fashion as we demonstrate in the inset of Fig. \ref{fig:plots}c). 
Note that the Dirac cones move closer to each other when the driving parameters are varied~\cite{Del13}, leading to a topological transition roughly around $\omega\approx \gamma$ when reducing $\omega$ for fixed $\mathcal{A}=0.2$. In the course of the transition, the direction of the edge state reverses, causing the opposite sign of the magnetization gradient observed in the inset. Even though the Dirac points move, the curves still collapse approximately away from a transition. Note that the low frequency regime supports additional edge states around quasienergy $\epsilon=\pi/ T$, which also feature a universal magnetization gradient following Eq. (\ref{eq:Medge}). Adiabaticity gives some control over the occupation of these edge states as well; the situation is more complex though. 

Experimentally the largest gaps can be achieved for very low frequencies ($\omega\sim 0.05\gamma$) - unfortunately these lie beyond reach of our simulation, due to the large system sizes required to provide enough energy resolution. Nevertheless we expect that the same arguments hold also in this regime, encouraged by the experimental observations we discuss in the next paragraph.

\emph{Experimental considerations.}
Experimentally, a sizable gap of $\Delta_0\sim 50 $meV has been induced by intense circularly-polarized laser pulses ($\hbar\omega=120$meV, $E_0\approx 2.5\!\times\! 10^7\,$V/m) in the Dirac-like spectrum of a TI surface~\cite{Ged13}.  The laser pulses hit an area of size $\sim 300 \mu$m and had a pulse length $250$fs containing $\sim10$ laser oscillations -- thus they were short enough ($\lesssim$ps) to avoid strong phonon relaxation~\cite{Hsi11,Sob12,Wan12}. Inspired by these values, consider an experiment with either graphene or a TI surface and a slightly higher light frequency or longer pulse duration such that sufficiently many laser cycles ($\sim 40$) are accommodated to realise both, a switch-on period of $5$ cycles and a sizeable time span of periodic driving.

The total mean current flowing along the edge is $I=e\Delta_0/h $ when the complete chiral edge state is filled across a gap $\Delta_0$. For a gap size of $\Delta_0\sim50$ meV as in Ref.~\cite{Ged13}, the current would thus be $I\approx 2 \mu$A. Let's imagine measuring the magnetic field produced by this current with a SQUID device of $r\sim\mu$m radius for temperatures around $4$K. Realistically the SQUID operates in a quantum limited regime with a magnetic field noise of $\sim 10^{-11}$T/$\sqrt{\rm Hz}$. In a region close to the current flow, the magnetic field would be $B\sim\mu_0 I/2 r\sim \mu$T. SQUID measurement times $\lesssim 1$ps would yield a signal to noise ratio of $\sim 0.1$. Present state of the art SQUIDS are not able to achieve these short time scales yet, but $\sim 10$ps seems to be within experimental reach ~\cite{Molweb}. In such a measurement, the observed mean magnetization is reduced by $1/10$ and the noise by $1/\sqrt{10}$, leading to a signal to noise ratio of $\sim0.03$. Careful design of the experiment and the laser parameters should be able to improve this ratio -- e.g. in graphene, the spin increases the current by a factor of $2$ and the higher Fermi velocity allows for larger gaps. Stacking of electrically isolated graphene layers would further enhance the field strength like in a coil and should be feasible technology-wise \cite{Pon13}. Given automated repetitive measurements we are thus expecting that the magnetic field can be probed. Note that operating the SQUID in the presence of an intense laser pulse poses a technological challenge itself. However, it will help that the pulses are very short and the fields circularly polarized.

Disorder should not be able to destroy the magnetization signature: first of all, an initial single Dirac cone and its preparation are stable against weak disorder, a situation found e.g.  on a TI surface. For the case of graphene, disorder-induced coupling of the two Dirac cones can in principle open a gap in the spectrum -  simulations however show that it requires extremely short-ranged disorder and weak disorder can ensure that the gap is very small. The Floquet phase itself is stable against weak disorder due to its topological nature.
Finally, band folding in combination with disorder scattering does provide extra decay channels for the edge states in the low frequency regime ~\cite{Zho11,Kun14}. However, these are suppressed due to their multi-photon nature and/or the extreme short-rangedness of disorder required. 

Similarly to the electron-phonon case, electron-electron interactions will generate apparent broadening of the one-electron spectral lines at least at the perturbative level; more sophisticated methods capture also quantum fluctuations in interacting systems~\cite{kamenev,Pol10}.  While the full effects of interactions are not calculable, it is believed that the edge states that give rise to the orbital magnetization signature are stable to interactions as they are protected by a topological invariant, although this line of argument has only been worked out fully in the equilibrium case.

Our claim that the discussed perturbations and decay mechanisms are unimportant for small enough pulse durations is supported by experimental observations from the MIT group of Nuh Gedik: the typical decay time scales found on a topological insulator surface are of the order of picoseconds \cite{Hsi11}, and the spectral density of states of the Floquet phase, including the gap, was found to be stable against these effects \cite{Ged13}.

To summarize we have investigated the mean magnetization of periodically-driven systems and proposed a protocol to realize and measure light-induced quantum Hall-like edge states, overcoming the typically fatal problems connected to Floquet phases. 

\emph{Acknowledgments.}
We thank  Alex Frenzel, Nuh Gedik, Fernando de Juan and Timm Rohwer for enriching discussions and John Clarke for his insights on the SQUID measurement constraints. Our research was supported by the Dutch Science Foundation NWO and the German Academic Exchange service DAAD (J.P.D.), as well as NSF DMR-1206515 (J.E.M.) and partially by Conacyt (B.M.F.). Computer resources were partially provided by NERSC.


\begin{thebibliography}{99}
\bibitem{footnote1}Note that for the inset of Fig. \ref{fig:Setup}, the electromagnetic gauge was adjusted to analyse the momentum structure.
\bibitem{footnote2}Larger $L$ do not modify the curves significantly.
\end{thebibliography}
\end{document}